\documentclass[twocolumn,showpacs,preprintnumbers,superscriptaddress,amsthm,amsmath,amssymb]{revtex4}
\usepackage{graphicx}
\usepackage{dcolumn}
\usepackage{bm}
\usepackage{color}
\usepackage{booktabs,longtable}
\usepackage{multirow}
\usepackage{CJK}
\usepackage{ltxtable}
\usepackage{rotating}
\usepackage[colorlinks,linkcolor=blue,anchorcolor=blue,citecolor=red]{hyperref}
\usepackage{setspace}

\begin{document}

\title{Continuum Skyrme Hartree-Fock-Bogoliubov theory with Green's function method\\
 for neutron-rich Ca, Ni, Zr, Sn isotopes}

\author{En-Bo Huo}
\affiliation{School of Physics and Microelectronics, Zhengzhou University,
Zhengzhou 450001, China}

\author{Ting-Ting Sun}
\email{ttsunphy@zzu.edu.cn}
\affiliation{School of Physics and
Microelectronics, Zhengzhou University, Zhengzhou 450001, China}

\author{Ke-Ran Li}
\affiliation{School of Physics and Microelectronics, Zhengzhou University,
Zhengzhou 450001, China}

\author{Xiao-Ying Qu}
\affiliation{School of Mechatronics Engineering, Guizhou Minzu University,
Guiyang 550025, China}

\author{Ying Zhang}
\affiliation{Department of Physics, School of Science, Tianjin University,
Tianjin 300072, China}

\date{\today}

\begin{abstract}

The possible exotic nuclear properties in the neutron-rich Ca, Ni, Zr, and
Sn isotopes are explored with the continuum Skyrme Hartree-Fock-Bogoliubov
theory formulated with the Green's function method, in which the pairing
correlation, the couplings with the continuum, and the blocking effects for
the unpaired nucleon in odd-$A$ nuclei are properly treated. In the $ph$
channel, the SLy4 parameter set is taken for the Skyrme interaction, and in
the $pp$ channel, the DDDI is adopted for the pairing interaction. With
those parameters, the available experimental two-neutron separation energies
$S_{\rm 2n}$ and one-neutron separation energies $S_{\rm n}$ are well
reproduced. Much shorter drip lines predicted by $S_{\rm n}$ are obtained
compared with those by $S_{\rm 2n}$. The systematic studies of the neutron
pairing energies $-E_{\rm pair}$ shown that values of the odd-$A$ nuclei are
much smaller in comparison with those of the neighboring even-even nuclei
due to the absent contribution of pairing energy by the unpaired odd
neutron. By investigating the single-particle structures, the rms radii and
the density distributions, the possible halo structures in the neutron-rich
Ca, Ni, and Sn isotopes are predicted, in which the sharp increases of rms
radii with significant deviations from the traditional $r\varpropto A^{1/3}$
rule and very diffuse spatial distributions in densities are observed. By
analyzing the contributions of different partial waves to the total neutron
density $\rho_{lj}(r)/\rho(r)$, the orbitals locating around the Fermi
surface especially those with low angular momenta are found the main reason
causing the extended nuclear density and large rms radii. Finally, the
numbers of the neutrons $N_{\lambda}$~($N_0$) occupied above the Fermi
surface $\lambda_n$~(in the continuum) are discussed, the behaviors of which
are basically consistent with those of pairing energy, supporting the key
role of the pairing correlations in the halo phenomena.

\end{abstract}

 \maketitle
\newcounter{mytempeqncnt}

\section{Introduction}
\label{Sec:Introduction}

The study of exotic nuclei far from the $\beta$ stability line is a very
challenging frontier topic in the nuclear physics
experimentally~\cite{Tanihata_PPNP_1995,Jonson_PhysRep_2004,Tanihata_PPNP_2013,Nakamura_AAPPSBL_2019}
and
theoretically~\cite{Mueller_ARNPS_1993,Hansen_ARNPS_1995,Casten_PPNP_2000,Jensen_RMP_2004,JMeng_PPNP_2006,Ershov_JPG_2010}.
The unstable nuclei with extreme $N/Z$ ratios, which are weakly bound systems,
have shown many exotic properties being different from the stable nuclei, such
as the halo
structures~\cite{Tanihata_PRL_1985,Minamisono_PRL_1992,Schwab_ZPA_1995,JMeng_PRL_1996,JMengJ_PRL_1998,SGZhou_PRC_2010,XXSun_SB_2021},
the changes of traditional magic
numbers~\cite{Ozawa_PRL_2000,Otsuka_PRL_2001,Rejmund_PRC_2007,Rosenbusch_PRL_2015,Chen_PRL_2019,XXSun_PLB_2018},
and new nuclear excitation modes~\cite{Zilges_PPNP_2005,Adrich_PRL_2005}, and
may also herald new physics~\cite{Zhou_POS_2017}. The study of exotic nuclei
is not only crucial for the fully understanding of the rich nuclear structures
and properties, the development of nuclear theory, but also vitally important
for the investigations of element synthesis and nuclear
astrophysics~\cite{Arnould_PhysRep_2007}. However, due to the extremely short
lifespan, the generation cross sections of the exotic nuclei are extremely
small, which makes much difficult to create them experimentally. For this
reason, more and more advanced large-scale radioactive beam facilities and
updated detector techniques have been built, upgraded, or planned in the
worldwide~\cite{JXia_NIMPRSA_2002,WLZhan_NPA_2010,Sturm_NPA_2010,Gales_NPA_2010,Motobayashi_NPA_2010,Thoennessen_NPA_2010,Zhou_NPR_2018}.
Besides, to overcome the difficulties of improving the current intensity and
quality of the radioactive nuclear beams, theoretical studies of exotic nuclei
provide an useful way for the layout of experiments and analysis of
experimental results.

The weakly bound nuclei are open quantum many-body systems, the theoretical
descriptions for which are very complicated. In the exotic nuclei, especially
the drip line nuclei, the neutron or proton Fermi surfaces are usually very
close to the continuum threshold. With the effects of the pairing correlation,
the valence nucleons have a certain probability to be scattered into the
continuum and occupy the resonance states therein, making the nuclear density
distribution very diffuse and extended. It is therefore essential to treat the
pairing correlations and the couplings with the continuum properly in the
theoretical descriptions of exotic
nuclei~\cite{Dobaczewski_PRC_1996,JMeng_NPA_1998,Grasso_PRC_2001}. Besides, in
the one-neutron halo nuclei, such as $^{31}$Ne~\cite{Nakamura_PRL_2009} and
$^{37}$Mg~\cite{Kobayashi_PRL_2014}, the blocking effect~\cite{Ring_book_2000}
should be considered to treat the unpaired odd nucleon.

The Hartree-Fock-Bogliubov~(HFB)~theory has achieved great successes in
describing exotic nuclei with an unified description of the mean field and the
pairing correlation by the Bogoliubov transformation~\cite{Ring_book_2000}.
Different models based on HFB theories have been taken to explore the exotic
nuclei, such as the Gogny-HFB theory~\cite{Decharg_PRC_1980}, the Skyrme-HFB
theory~\cite{Dobaczewski_NPA_1984}, the relativistic continuum
Hartree-Bogoliubov~(RCHB)~theory~\cite{JMeng_NPA_1998}, and the density
dependent relativistic
Hartree-Fock-Bogoliubov~(RHFB)~theory~\cite{WHLong_PRC_2010}. Besides, to
explore the halo phenomena in deformed nuclei, these models have been extended
to the deformed framework, such as the deformed relativistic
Hartree-Bogoliubov~(DRHB)~theory~\cite{SGZhou_PRC_2010,LLLi_PRC_2012,YChen_PRC_2012}
and the coordinate-space Skyrme-HFB
approach~\cite{JCPei_PRC_2013,YZhang_PRC_2013,JCPei_PRC_2014,YShi_PRC_2018}.

Traditionally, these~H(F)B~equations are often solved in configuration spaces,
that is the basis expansion method~\cite{Gambhir_AP_1990}. However, the
calculations are strongly related to the space size and the shape of the
expanded basis. For the harmonic oscillator basis with a tail in a Gaussian
function shape which decays fast at large distances, although it is very
efficient to describe the stable nuclei, great difficulties have been met for
the description of the exotic nuclei with extended density distributions. To
avoid these problems, basises with proper shape, such as the Woods-Saxon
basis~\cite{SGZhou_PRC_2003} and the transformed harmonic oscillator
basis~\cite{Stoitsov_PRC_1998,Stoitsov_CPC_2005}, have been applied for the
exotic nuclei. For example, to explore the halo phenomena in deformed
nuclei~\cite{KYZhang_PRC_2020,Kim_PRC_2022}, the DRHB theory based on a
Woods-Saxon basis~\cite{SGZhou_PRC_2010,LLLi_PRC_2012} has been developed. In
contrast with the basis expansion method, solving the HFB equation in the
coordinate space is believed to be simply effective, where the Nomerov or
Runge-Kutta method is often applied. In the coordinate space, discretized
method is often taken to solve the HFB equations with the box boundary
condition, where a series of discrete levels could be obtained easily.
Although this approach is widely used, some flaws have been found, such as the
nonphysical drop of the nuclear densities at the box boundary and the continua
and the resonance states therein are discretized. In contrast, the Green's
function~(GF)~method~\cite{Tamura_PRB_1992,Foulis_PRA_2004,Economou_book_2006}
in the coordinate space can avoid these problems and owns great advantages,
which can describe the asymptotic behaviors of wave functions properly,
provide the energies and widths of the resonant states directly, and treat the
bound states and the continua on the same footing.

Due to the great advantages, the Green's function method has been applied
extensively in nuclear physics to study the contribution of the continuum to
the nuclear structures and excitations. As early as in 1987, Belyaev \emph{et
al.} has constructed the Green's function for the HFB
equation~\cite{Belyaev_SJNP_1987}. Afterwards, this HFB Green's function was
applied to the quasiparticle random-phase
approximation~(QRPA)~\cite{Matsuo_NPA_2001}, which was further used to
describe the collective excitations coupled to the
continuum~\cite{Matsuo_PTPS_2002,Matsuo_PRC_2005,Matsuo_PRC_2010,Shimoyama_PRC_2011,Shimoyama_PRC_2013,Matsuo_PRC_2015}.
In 2009, the continuum HFB theory in a coupled channel representation was
developed to explore the effects of the continuum and pairing correlation in
deformed Mg isotopes near the neutron drip line~\cite{Oba_PRC_2009}. In 2011,
Zhang \emph{et al.} developed the fully self-consistent continuum Skyrme-HFB
theory with Green's function method~\cite{YZhang_PRC_2011}, which was further
applied to investigate the giant halos~\cite{YZhang_PRC_2012} and the effects
of pairing correlation on the quasiparticle
resonances~\cite{Qu_SciChina_2019,YZhang_PRC_2020}. In 2019, to explore the
halo phenomena in the neutron-rich odd-$A$ nuclei, the self-consistent
continuum Skyrme-HFB theory was further extended by including the blocking
effect~\cite{TTSun_PRC_2019}. In recent years, considering the great successes
of the covariant density functional theory~(CDFT) in the studies of nuclear
structure~\cite{JMeng_JPG_2015,TTSun_PRC_2016,WLLV_JPG_2017,TTSun_PRC_2017,TTSun_CPC_2018,ZXLiu_PRC_2018,TTSun_PRD_2019,CChen_2021_SCPMA,Tanimura_PRC_2022},
Green's function method has also been combined with the CDFT. Introducing the
Green's function method to the relativistic mean filed theory~(GF-RMF), the
single-particle level structures including the bound states and resonant
states as well as the related physical researches such as the pseudospin
symmetry are investigated for
neutrons~\cite{TTSun_PRC_2014,TTSun_PRC_2019_034310},
protons~\cite{TTSun_JPG_2016}, and $\Lambda$ hyperons~\cite{SHRen_PRC_2017}.
Besides, it has been confirmed that no matter for the broad or narrow resonant
states, exact values of the energies and widths could be obtained by searching
for the poles of Green's function or the extremes of the density of
states~\cite{Chen_CPC_2020,Wang_NST_2021}. By combining the Green's function
method with the RCHB theory, the pairing correlation and continuum are well
described in the giant halos of the Zr isotopes~\cite{TTSun_Sci_2016}. By
extending the GF-RMF model to the coupled channel representation, the halo
candidate nucleus $^{37}$Mg reported experimentally is analyzed and confirmed
to be a $p$-wave one-neutron halo according to the Nilsson
levels~\cite{TTSun_PRC_2020}. All those works have proved the great successes
of Green's function method in the descriptions of the continuum.

In this paper, the neutron-rich Ca, Ni, Zr, and Sn isotopes are investigated
systematically taking the continuum Skyrme-HFB theory formulated with the
Green's function method in the coordinate space, which has treated the pairing
correlations, the couplings with the continuum, and the blocking effect for
the odd unpaired nucleon properly. The paper is organized as follows: In
Sec.~\ref{Sec:Theory}, we introduce briefly the continuum Skyrme-HFB theory.
Numerical details are presented in Sec.~\ref{Sec:Numerical details}. After
giving the results and discussions in Sec.~\ref{Sec:Results}, finally
conclusions are drawn in Sec.~\ref{Sec:Summary}.

\section{THEORETICAL FRAMEWORK}
\label{Sec:Theory}

In the Hartree-Fock-Bogoliubov~(HFB)~theory~\cite{Ring_book_2000}, the pair
correlated nuclear system is described in terms of independent quasiparticles
by the Bogoliubov transformation. The HFB equation in the coordinate
space~\cite{Dobaczewski_PRC_1996} is,
\begin{equation}
\left(
     \begin{array}{cc}
       h-\lambda    & \tilde{h}      \\
      \tilde{h}^{*} & -h^{*}+\lambda \\
     \end{array}
\right)
\phi_{i}({\bm r}\sigma)
=E_{i}
\phi_{i}({\bm r}\sigma),
\label{EQ:HFB}
\end{equation}
where $E_{i}$ is the quasiparticle energy, $\phi_i({\bm r}\sigma)$ is the
quasiparticle wave function, and $\lambda$ is the Fermi energy determined by
constraining the expectation value of the nucleon number. The HF Hamiltonian
$h(\bm{r}\sigma, \bm{r'}\sigma')$ and the pair Hamiltonian
$\tilde{h}(\bm{r}\sigma, \bm{r'}\sigma')$ are obtained by the variation of the
total energy functional with respect to the particle density
$\rho(\bm{r}\sigma, \bm{r}'\sigma')$ and the pair density
$\tilde{\rho}(\bm{r}\sigma, \bm{r}'\sigma')$, respectively. The solutions of
the HFB equations have two symmetric branches, one of which is positive
($E_i>0$) with the quasiparticle wave function $\phi_i({\bm r}\sigma)$ while
the other one is negative ($-E_i<0$) with the conjugate wave function
$\bar{\phi}_{\tilde{i}}({\bm r}\sigma)$. The wave functions $\phi_i({\bm
r}\sigma)$ and $\bar{\phi}_{\tilde{i}}({\bm r}\sigma)$ are in two components
and they are related as,
\begin{equation}
\phi_i({\bm r}\sigma)=
\left(
  \begin{array}{c}
    \varphi_{1,i}({\bm r}\sigma) \\
    \varphi_{2,i}({\bm r}\sigma) \\
  \end{array}
\right),~~
\bar{\phi}_{\tilde{i}}({\bm r}\sigma)=
\left(
  \begin{array}{r}
    -\varphi_{2,i}^*({\bm r}\tilde{\sigma}) \\
    \varphi_{1,i}^*({\bm r}\tilde{\sigma}) \\
  \end{array}
\right),
\label{EQ:HFBwf}
\end{equation}
with $\varphi({\bm r}\tilde{\sigma})=-2\sigma\varphi({\bm r},-\sigma)$. Here
and hereafter, we follow the notations in Ref.~\cite{Matsuo_NPA_2001} for
convenience.

For an even-even nucleus, the ground state $|\Phi_{0}\rangle$ is a
quasiparticle vacuum and the particle density $\rho(\bm{r}\sigma, \bm{r}'\sigma')$ and pairing density $\tilde{\rho}(\bm{r}\sigma, \bm{r}'\sigma')$ are written as%
\begin{subequations}%
\begin{eqnarray}%
&&\rho(\bm{r}\sigma, \bm{r}'\sigma')\equiv \langle\Phi_{0}|c^{\dag}_{\bm{r}'\sigma'}c_{\bm{r}\sigma}|\Phi_{0}\rangle,\\%
&&\tilde{\rho}(\bm{r}\sigma, \bm{r}'\sigma')\equiv \langle\Phi_{0}|c_{\bm{r}'\tilde{\sigma}'}c_{\bm{r}\sigma}|\Phi_{0}\rangle,%
\end{eqnarray}%
\label{EQ:rho}%
\end{subequations}%
where $c_{\bm{r}\sigma}^{\dag}$ and $c_{\bm{r}\sigma}$ are the particle
creation and annihilation operators, respectively. The densities can be
unified as a generalized density matrix $R({\bm r}\sigma,{\bm r}'\sigma')$,
with $\rho(\bm{r}\sigma, \bm{r}'\sigma')$ and $\tilde{\rho}(\bm{r}\sigma,
\bm{r}'\sigma')$ being the ``11'' and ``22'' element, respectively. With the
quasiparticle wave functions, $R({\bm r}\sigma,{\bm r}'\sigma')$ can be
written in a simple form,
\begin{equation}
R({\bm r}\sigma,{\bm r}'\sigma')=
\sum_{i}\bar{\phi}_{\tilde{i}}({\bm r}\sigma)\bar{\phi}_{\tilde{i}}^{\dag}({\bm r'}\sigma').
\label{EQ:Rmatrix}
\end{equation}

For an odd-$A$ nucleus, the last odd nucleon is unpaired, for which the
blocking effect should be considered. The nuclear ground state in this case is
a one-quasiparticle state $|\Phi_{1}\rangle$, which can be constructed based
on a HFB vacuum $|\Phi_{0}\rangle$ as
\begin{equation}
 |\Phi_{1}\rangle=\beta_{i_{\rm b}}^{\dag}|\Phi_{0}\rangle,
\end{equation}
where $\beta_{i_{\rm b}}^{\dag}$ is the quasiparticle creation operator and
$i_{\rm b}$ denotes the blocked quasiparticle level occupied by the odd
nucleon. Accordingly, the particle density $\rho(\bm{r}\sigma,\bm{r}'\sigma')$
and the pairing density $\tilde{\rho}(\bm{r}\sigma, \bm{r}'\sigma')$ are%
\begin{subequations}%
\begin{eqnarray}%
&&\rho(\bm{r}\sigma, \bm{r}'\sigma')\equiv \langle\Phi_{1}|c^{\dag}_{\bm{r}'\sigma'}c_{\bm{r}\sigma}|\Phi_{1}\rangle,\\%
&&\tilde{\rho}(\bm{r}\sigma, \bm{r}'\sigma')\equiv \langle\Phi_{1}|c_{\bm{r}'\tilde{\sigma}'}c_{\bm{r}\sigma}|\Phi_{1}\rangle,%
\end{eqnarray}%
\label{EQ:rho_odd}%
\end{subequations}%
and the generalized density matrix $R({\bm r}\sigma,{\bm r}'\sigma')$ becomes
\begin{eqnarray}
R({\bm r}\sigma,{\bm r}'\sigma')&=&
\sum_{i:{\rm all}}\bar{\phi}_{\tilde{i}}({\bm r}\sigma)\bar{\phi}_{\tilde{i}}^{\dag}({\bm r'}\sigma')
\label{EQ:Rmatrix_odd}\\
&&-\bar{\phi}_{\tilde{i}_{\rm b}}({\bm r}\sigma)\bar{\phi}_{\tilde{i}_{\rm b}}^{\dag}({\bm r'}\sigma')
+\phi_{i_{\rm b}}({\bm r}\sigma)\phi_{i_{\rm b}}^{\dag}({\bm r'}\sigma'),\nonumber
\end{eqnarray}
where two more terms are introduced compared with those for the even-even
nuclei.

In the conventional Skyrme-HFB theory, the HFB equation~(\ref{EQ:HFB}) in the
coordinate space is often solved with the box boundary condition, and a series
of the discretized eigensolutions including the quasiparticle energy $E_i$ and
the corresponding wave functions $\phi_i({\bm r}\sigma)$ could be obtained.
Then the generalized density matrix $R({\bm r}\sigma,{\bm r}'\sigma')$ can be
calculated by summing those discretized quasiparticle states as shown in
Eqs.~(\ref{EQ:Rmatrix}) and (\ref{EQ:Rmatrix_odd}). We call this method the
box-discretized approach. However, the applicability of the box boundary
condition in the description of exotic nuclei especially those close to the
drip line is not good. A large enough coordinate space (or box size) should be
taken to describe the very extended density distribution.

Green's function method can avoid these problems of box-discretized approach,
which impose the correct asymptotic behaviors on the wave functions,
especially for the weakly-bound sates and the continuum, and calculate the
densities. The Green's function $G({\bm r}\sigma,{\bm r}'\sigma';E)$ with an
arbitrary quasiparticle energy $E$ defined for the coordinate-space HFB
equation obeys,
\begin{eqnarray}
&&\left[E-\left(
     \begin{array}{cc}
       h-\lambda    & \tilde{h}      \\
      \tilde{h}^{*} & -h^{*}+\lambda \\
     \end{array}
\right)\right]
G({\bm r}\sigma,{\bm r}'\sigma';E)\nonumber\\
&&~~~~=\delta({\bm r}-{\bm r'})\delta_{\sigma\sigma'},
\end{eqnarray}
which is a $2\times 2$ matrix. The generalized density matrix $R({\bm
r}\sigma,{\bm r}'\sigma')$ in Eq.~(\ref{EQ:Rmatrix_odd}) can be calculated by
the integrals of the Green's function performed on the complex quasiparticle
energy plane as
\begin{eqnarray}
R({\bm r}\sigma,{\bm r}'\sigma')&=&\frac{1}{2\pi i}\left[\oint_{C_{E<0}}dE G({\bm r}\sigma,{\bm r}'\sigma';E)\right.\nonumber\\
&&-\oint_{C_{\rm b}^-}dE G({\bm r}\sigma,{\bm r}'\sigma';E)\nonumber\\
&&\left.+\oint_{C_{\rm b}^+}dE G({\bm r}\sigma,{\bm r}'\sigma';E)\right],
\label{Rmax_GF_odd}
\end{eqnarray}
where the contour path $C_{E<0}$ encloses all the negative quasiparticle
energies $-E_i<0$, $C_{\rm b}^-$ encloses only the pole of $-E_{i_{\rm b}}$,
and $C_{\rm b}^+$ encloses only the pole of $E_{i_{\rm b}}$.

In the spherical case, the quasiparticle wave functions $\phi_{i}({\bm
r}\sigma)$ and $\bar{\phi}_{\tilde{i}}({\bm r}\sigma)$ are only dependent on
the radial parts, and they can be expanded as
\begin{subequations}
\begin{eqnarray}
&&\phi_i({\bm r}\sigma)=\frac{1}{r}\phi_{nlj}(r)Y_{jm}^l(\hat{\bm r}\sigma),\\
&&\bar{\phi}_{\tilde{i}}({\bm r}\sigma)=\frac{1}{r}\bar{\phi}_{nlj}(r)Y_{jm}^{l*}(\hat{\bm r}\tilde{\sigma}),\\
 &&\phi_{nlj}(r)= \left(
  \begin{array}{c}
    \varphi_{1,nlj}(r) \\
    \varphi_{2,nlj}(r) \\
  \end{array}
\right),
 \bar{\phi}_{nlj}(r)= \left(
  \begin{array}{c}
    -\varphi_{2,nlj}^{*}(r) \\
    ~~~\varphi_{1,nlj}^*(r) \\
  \end{array}
\right),~~~~~~
\end{eqnarray}
\end{subequations}%
where $Y_{jm}^l(\hat{\bm r}\sigma)$ is the spin spherical harmonic, and
$Y_{jm}^l(\hat{\bm r}\tilde{\sigma})=-2\sigma Y_{jm}^l(\hat{\bm r}-\sigma)$.
Similarly, the generalized density matrix $R({\bm r}\sigma,{\bm r}'\sigma')$
and the Green's function $G({\bm r}\sigma,{\bm r}'\sigma';E)$ can be expanded
as
\begin{subequations}%
\begin{eqnarray}
&& R({\bm r}\sigma, {\bm r'}\sigma')=\sum_{ljm}Y_{jm}^l(\hat{\bm r}\sigma)R_{lj}(r,r')Y_{jm}^{l*}(\hat{\bm r}'\sigma'),\\
&&G({\bm r}\sigma, {\bm r'}\sigma';E)=\sum_{ljm}Y_{jm}^l(\hat{\bm r}\sigma)\frac{\mathcal{G}_{lj}(r,r';E)}{rr'}Y_{jm}^{l*}(\hat{\bm r}'\sigma'),~~~~~~%
\end{eqnarray}%
\end{subequations}%
where $R_{lj}(r,r')$ and $\mathcal{G}_{lj}(r,r';E)$ are the radial parts of
the generalized density matrix and Green's function, respectively.

As a result, the radial local generalized density matrix $R(r)=R(r,r)$ can be
expressed by the radial HFB Green's function $\mathcal{G}_{lj}(r,r';E)$ as
\begin{eqnarray}
R(r)&=&\sum_{lj}R_{lj}(r,r)=\frac{1}{4\pi r^{2}}\left[\sum_{lj:{\rm all}}(2j+1) \sum_{n:{\rm all}} \bar{\phi}_{nlj}^{2}(r)\right.\nonumber\\
&&\left.-\bar{\phi}_{n_{\rm b} l_{\rm b} j_{\rm b}}^{2}(r)+\phi_{n_{\rm b} l_{\rm b} j_{\rm b}}^{2}(r)\right] \nonumber\\
&=& \frac{1}{4\pi r^{2}} \frac{1}{2\pi i}\left[\sum_{lj:{\rm all}}(2j+1) \oint_{C_{E<0}} dE \mathcal{G}_{lj}(r,r;E)\right.\nonumber\\
&&\left.-\oint_{C_{\rm b}^{-}} dE \mathcal{G}_{l_{\rm b} j_{\rm b}}(r,r;E)+\oint_{C_{\rm b}^{+}} dE \mathcal{G}_{l_{\rm b} j_{\rm b}}(r,r;E)\right].\nonumber\\
\end{eqnarray}
From the radial generalized matrix $R(r)$, one can easily obtain the radial
local particle density $\rho(r)$ and pair density $\tilde{\rho}(r)$, which are
the ``11'' and ``12'' components of $R(r)$, respectively. In the same way, one
can express other radial local densities needed in the functional of the
Skyrme interaction, such as the kinetic-energy density $\tau(r)$, the
spin-orbit density $J(r)$, etc., in terms of the radial Green's function. For
the construction of the Green's function, see
Refs.~\cite{YZhang_PRC_2011,TTSun_PRC_2019}.

\begin{figure*}[ht!]
\includegraphics[width=0.8\textwidth]{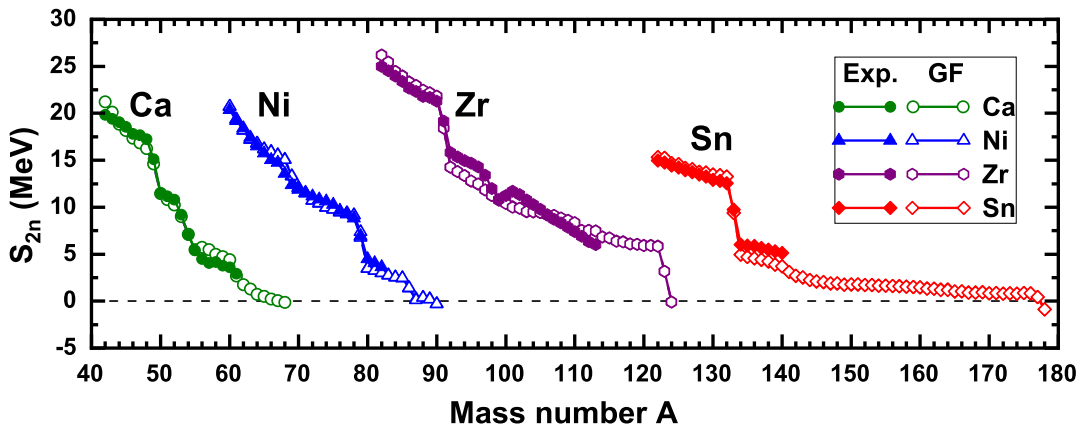}
\caption{(Color online) Two-neutron separation energies $S_{\rm 2n}$ in the
Ca, Ni, Zr, and Sn isotopes as a function of the mass number $A$. Open symbols
represent the results calculated by the continuum Skyrme HFB theory with the
SLy4 parameter set while the solid ones are the experimental
data~\cite{MWang_CPC_2021}.}
\label{Fig1}
\end{figure*}

\section{Numerical details}
\label{Sec:Numerical details}

For the Skyrme interaction in the $ph$ channel, the SLy4 parameter
set~\cite{Chabanat_NPA_1998} is adopted. For the pairing interaction in the
$pp$ channel, a density dependent $\delta$ interaction (DDDI) is taken,
\begin{equation}
v_{\rm pair}({\bm r},{\bm r'})=\frac{1}{2}(1-P_{\sigma})V_0\left[1-\eta\left(\frac{\rho({\bm r})}{\rho_0}\right)^{\alpha}\right]\delta({\bm r}-{\bm r'}),
\label{EQ:DDDI}
\end{equation}
with which the pair Hamiltonian $\tilde{h}({\bm r}\sigma, {\bm r}'\sigma')$ is
reduced to a local pair potential~\cite{Dobaczewski_NPA_1984},
\begin{equation}
 \Delta({\bm r})=\frac{1}{2}V_0\left[1-\eta\left(\frac{\rho({\bm r})}{\rho_0}\right)^\alpha\right]\tilde{\rho}({\bm r}),
\end{equation}
where the strength of the pairing force $V_{0}=-458.4$~MeV$\cdot$fm$^{3}$, the
density $\rho_{0}=0.08$~fm$^{-3}$, and other parameters $\eta=0.71$,
$\alpha=0.59$, which are constrained by reproducing the experimental neutron
pairing gaps for the Sn
isotopes~\cite{Matsuo_PRC_2006,Matsuo_NPA_2007,Matsuo_PRC_2010}. Besides, with
these parameters, DDDI can reproduce the scattering length $a=-18.5$~fm in the
$^{1}S$ channel of the bare nuclear force in the low density
limit~\cite{Matsuo_PRC_2006}. The cut-off of the quasiparticle states are
taken with maximal angular momentum $j_{\rm max}=25/2$ and the maximal
quasiparticle energy $E_{\rm cut}=60$~MeV.

The HFB equation is solved in the coordinate space with the space size $R_{\rm
box}=20$~fm and mesh size $dr=0.1$~fm. To calculate the densities with the
Green's function, the integrals of the Green's functions are performed along a
contour path $C_{E<0}$, which is chosen as a rectangle with the height $\gamma
= 0.1$~MeV and the length $E_{\rm cut} = 60$~MeV to enclose all the
quasiparticle states with negative energies. For the odd-$A$ nuclei, two more
contour paths $C_{\rm b}^+$ and $C_{\rm b}^-$, which only enclose the blocked
quasiparticle states at energies $E_{i_{\rm b}}$ and $-E_{i_{\rm b}}$, are
introduced due to the blocking effect of the odd unpaired nucleon. For these
details, it can be referred to Ref.~\cite{TTSun_PRC_2019}. To do the contour
integration, an energy step of $\Delta E = 0.01$~MeV on the contour path is
adopted.

\begin{figure}[h!]
\includegraphics[width=0.48\textwidth]{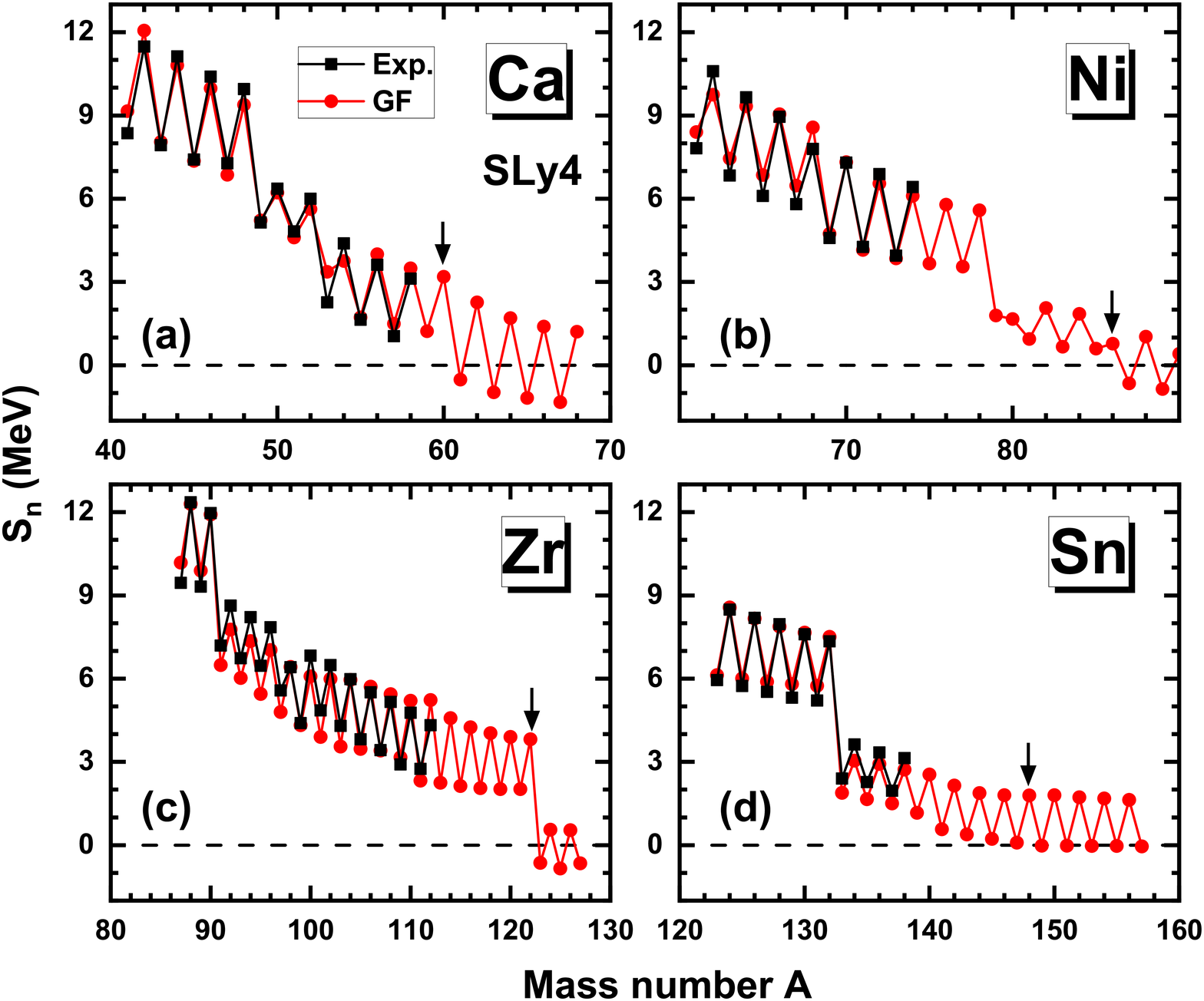}
\caption{ (Color online) Single-neutron separation energies $S_{\rm n}$ as a
function of the mass number $A$ in the (a)~Ca, (b)~Ni, (c)~Zr, and (d)~Sn
isotopes calculated by the continuum Skyrme-HFB theory with the SLy4 parameter
set~(red circles), in comparison with the experimental
data~\cite{MWang_CPC_2021}~(black squares).}
\label{Fig2}
\end{figure}

\section{RESULTS AND DISCUSSION}\label{Sec:Results}

In Fig.~\ref{Fig1}, the two-neutron separation energies $S_{\rm 2n}(N,Z) =
E(N-2,Z) - E(N, Z)$ are plotted for the even-even and odd-even Ca, Ni, Zr, and
Sn isotopes. Open symbols are those calculated by the continuum Skyrme-HFB
theory with the SLy4 parameter set, in comparison with the available
experimental data~\cite{MWang_CPC_2021} denoted by the solid symbols. Good
agreements with the experimental data could be observed, indicating the
reliability of the continuum Skyrme-HFB theory in the prediction of the
neutron drip line. The traditional shell closures, i.e., $N=28$ in the Ca
isotopes, $N=50$ in the Ni isotopes, $N=50, 82$ in the Zr isotopes, and $N=82$
in the Sn isotopes, could also be observed, where the $S_{\rm 2n}$ drops
sharply. For example, in the Sn chain, $S_{\rm 2n}$ drops from $13.25$~MeV at
$^{132}$Sn to $4.94$~MeV at $^{134}$Sn with the neutron number exceeding the
magic number $N=82$. In the Ca, Ni, and Zr chains, the two-neutron separation
energies quickly reach zero at large mass range, resulting in relative short
neutron drip lines, which are $^{67}$Ca, $^{89}$Ni, and $^{123}$Zr,
respectively. On the contrary, in the Sn chain, $S_{\rm 2n}$ keeps less than
$1.0$~MeV in a wide mass region after the gap of $N=82$ and finally becomes
negative until $A=178$, determining $^{177}$Sn as the neutron drip line
nucleus. Those weakly bound nuclei are quite interesting due to possible
appearance of neutron halos although it is very hard to reach experimentally.
In addition, the exploration of the neutron drip line and the determination of
the limit of the nuclear existence are significantly important in nuclear
physics. However, various theoretical studies show that the predicated neutron
drip line is very model dependent~\cite{XWXia_2020_ADNDT}. Moreover, different
physical quantities or criteria will also predict very different neutron drip
lines.

To explore the neutron drip lines in the Ca, Ni, Zr, and Sn isotopes, in
Fig.~\ref{Fig2}, the single-neutron separation energies $S_{\rm
n}(N,Z)=E(N-1,Z)-E(N,Z)$ are also plotted. Results by the continuum Skyrme-HFB
theory with the SLy4 parameter set are denoted by red circles, which are
consistent very well with the experimental data~\cite{MWang_CPC_2021} denoted
by the black squares. Strong odd-even staggering is observed in all isotopes.
In general, the $S_{\rm n}$ in the even-even nucleus is around $2\sim3$~MeV
larger than the neighbouring odd-$A$ nuclei, which could be explained by that
the unpaired odd neutron contributes no pairing energy. As a result, compared
with those in Fig.~\ref{Fig1}, the neutron drip lines determined by the
one-neutron separation energy are greatly shortened. In the Ca, Ni, Zr, and Sn
isotopes, the drip line nuclei are $^{60}$Ca, $^{86}$Ni, $^{122}$Zr, and
$^{148}$Sn, respectively, the positions of which are indicated by the black
arrows. Outside the neutron drip line determined by $S_{\rm n}$, the bound
even-even nuclei behave as the interesting Borromean systems. For example,
based on the bound nucleus $^{60}$Ca, $^{60}$Ca$+n$ is unbound while
$^{60}$Ca$+n+n$ is bound.

\begin{figure}[ht!]
\includegraphics[width=0.45\textwidth]{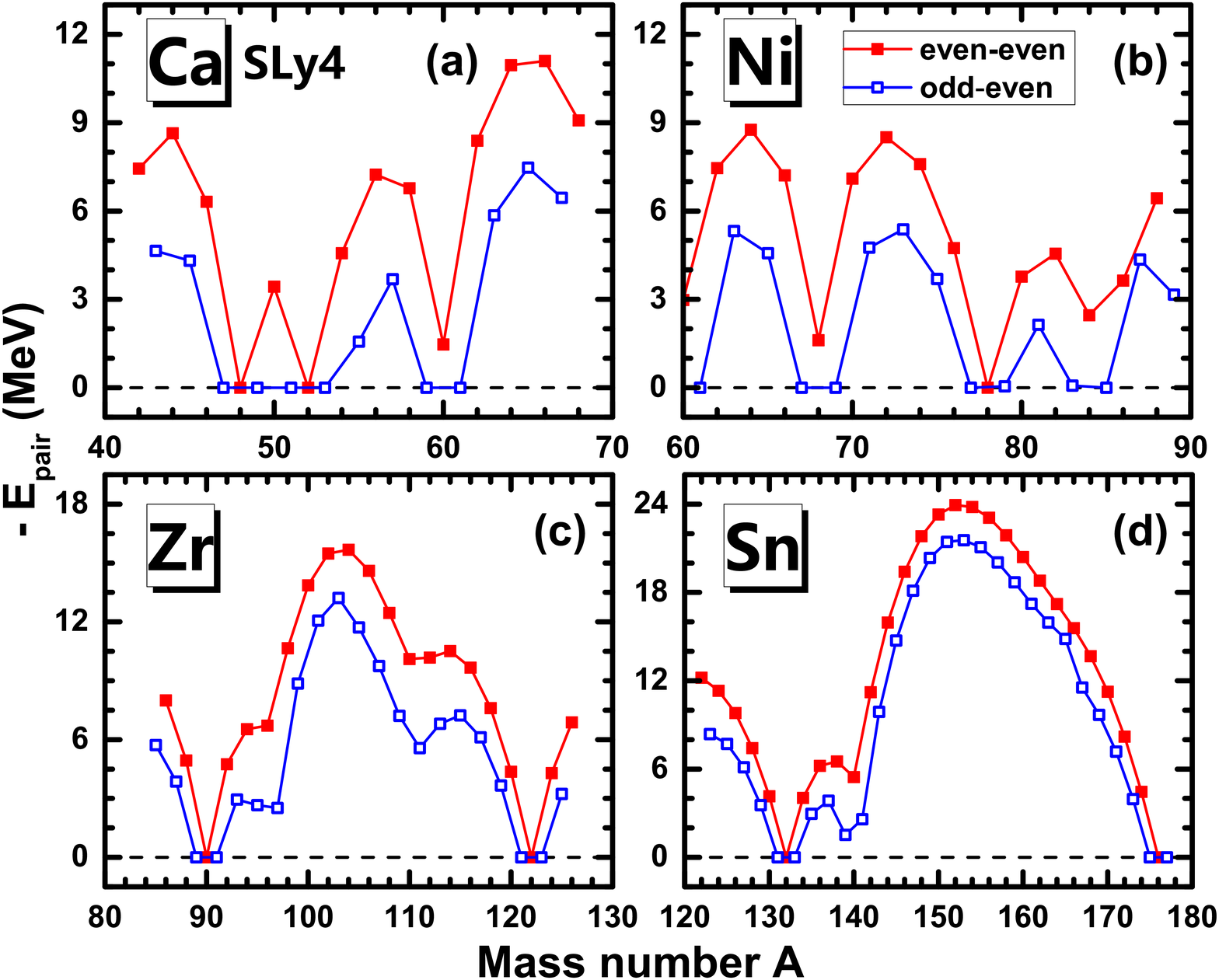}
\caption{(Color online) Neutron pairing energy $-E_{\rm pair}$ as a function of the mass number $A$ in the (a)~Ca, (b)~Ni, (c)~Zr, and (d)~Sn isotopes calculated by the continuum Skyrme-HFB theory with the SLy4 parameter set. The filled and open squares represent the results of the even-even and odd-even nuclei, respectively.}
\label{Fig3}%
\end{figure}%

In Fig.~\ref{Fig3}, we plot the neutron pairing energy $E_{\rm pair}$, which
is expressed as,
\begin{equation}
E_{\rm pair}=\frac{1}{2}\int d{\bm r}\Delta({\bm r})\tilde{\rho}({\bm r}).
\end{equation}
The red solid symbols are for the even-even nuclei and open symbols for the
odd-$A$ nuclei. The neutron pairing energies $E_{\rm pair}$ for the odd-$A$
nuclei are much smaller than those of the neighboring even-even nuclei, due to
the absent contribution of pairing energy by the unpaired neutron. This also
explains why the drip line determined by the single-neutron separation energy
$S_{\rm n}$ is much shorter than that obtained by the two-neutron separation
energy $S_{\rm 2n}$. In addition, the pairing energy $E_{\rm pair}$ in each
panel behaves as a wave packet as a function of the nuclear mass number, with
the wave trough locating at the neutron magic numbers and the crest of wave
appearing when the neutron number reaching the half full of the shell. For
example, in the Sn isotopes, a large wave package starting from $N=82$ and
ending at $N=126$ could be observed, with the highest value at the position of
the neutron number $N=102$. As a result, the traditional shell closures, i.e.,
$N=28, 40$ in the Ca isotopes, $N=40, 50$ in the Ni isotopes, $N=50, 82$ in
the Zr isotopes, and $N=82,126$ in the Sn isotopes, could also be obviously
observed, which are consistent with those obtained in Fig.~\ref{Fig1}. In
addition, a sub-shell $N=32$ is also observed in the Ca isotopes.

In the following, we will explore the possible neutron halo phenomena in the
Ca, Ni, Zr, and Sn isotopes, especially in the weakly bound nuclei near the
drip line, in which the neutron Fermi surfaces are very close to the continuum
threshold and the valence neutron could be easily scattered to the continuum
and occupy the resonant states.

\begin{figure}[t!]
\includegraphics[width=0.45\textwidth]{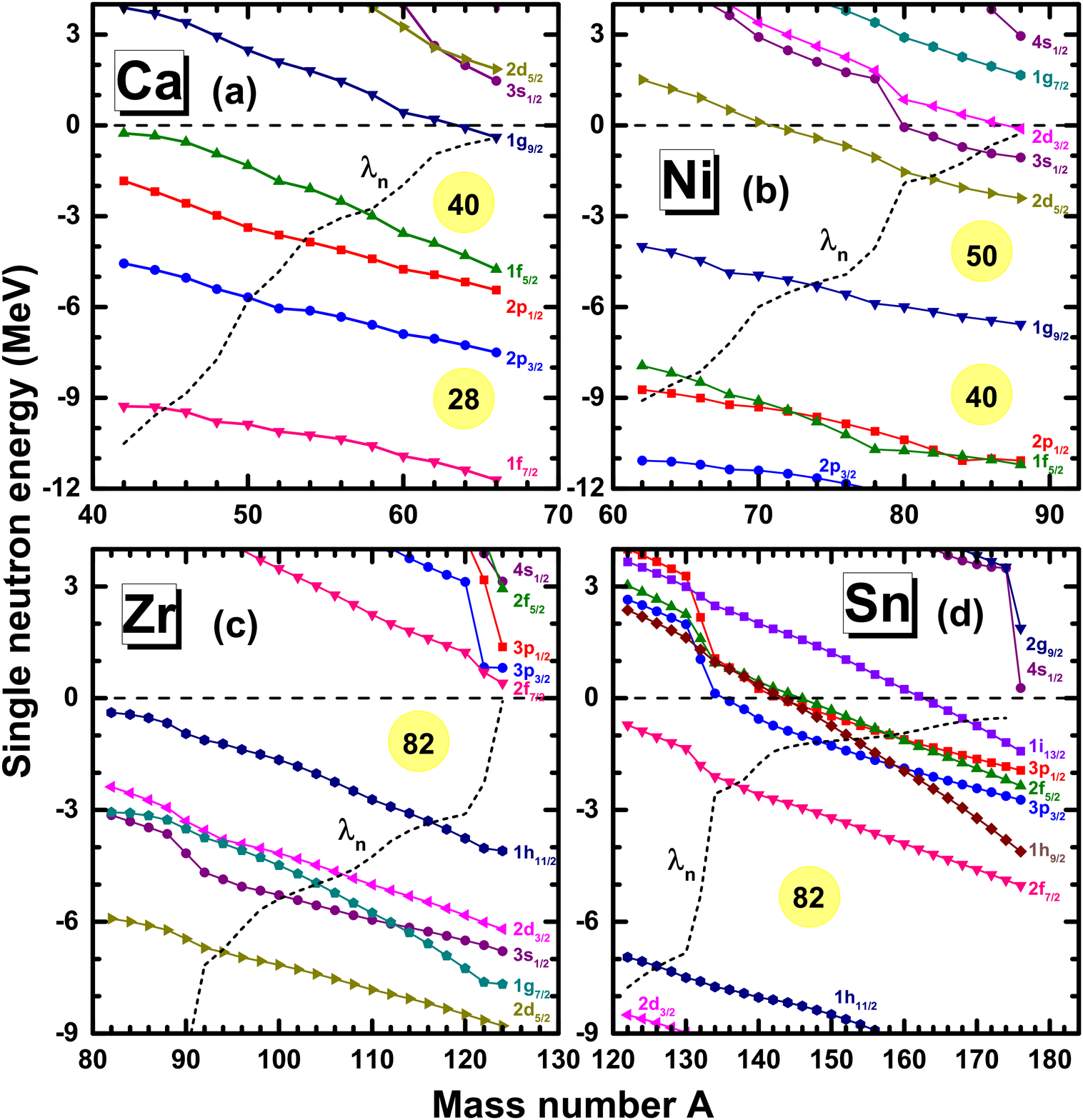}
\caption{(Color online) Canonical single-particle levels for neutrons around the fermi
surface in the (a)~Ca, (b)~Ni, (c)~Zr, and (d)~Sn isotopes calculated by the
continuum Skyrme-HFB theory. The short dashed lines represent the Fermi
surface.}
\label{Fig4}
\end{figure}

In Fig.~\ref{Fig4}, the neutron canonical single-particle structure as
functions of the mass number are plotted for the (a)~Ca, (b)~Ni, (c)~Zr, and
(d)~Sn isotopes. For the details how to get the canonical single-particle
levels, it can be referred to Refs.~\cite{XYQu_PRC_2019,XYQu_PRC_2022}. The
neutron Fermi energy $\lambda_{\rm n}$ as well as the canonical
single-particle energies $\varepsilon$ are shown. With the neutron number
increasing, the Fermi energy $\lambda_{\rm n}$ in each chain is raised up and
finally reaches the continuum threshold while all the HF single-particle
levels fall down. The traditional shell closures, i.e., $N=28, 40$ in the Ca
isotopes, $N=40, 50$ in the Ni isotopes, $N=82$ in the Zr isotopes, and $N=82$
in the Sn isotopes, could be observed, where the big gaps could be observed
obviously. Different single-particle structures are revealed in the Ca, Ni,
Zr, and Sn chains, based on which halos could be formed easily or hardly. In
the Ni chain, above the shell closure of $N=50$, there are several weakly
bound states and the low-lying resonant sates with small angular momenta,
i.e., $2d_{5/2}$, $3s_{1/2}$, and $2d_{3/2}$, which create good opportunities
for the formation of halos. For the Sn chain, which is quite similar to the Ni
chain but has more advantages for halo formation, more weakly bound states and
the low-lying resonant sates exist above the $N=82$ shell closure. Besides,
the Fermi surface $\lambda_{\rm n}$ approaches zero gradually and the Sn
isotopes in a large mass reagin are weakly bound. In the Ca chain, above the
shell closure of $N=40$, there is mainly the 1$g_{9/2}$ state, which evolves
from a single-particle resonant state ($A \leq 62$) to a weakly bound level
($A \geq 64$). Although there is a visitable possibility for valence neutrons
occupy 1$g_{9/2}$ orbital, the contributed density is very localized due to
the big central barrier. In the neutron-rich Ca isotopes, the low-lying
resonant states $3s_{1/2}$ and $2d_{5/2}$ also play important roles. The worst
case happens in the Zr chain where the shell closure of $N=82$ locates around
the threshold of continuum and it's hard for the valence neutrons to overcome
the big gap and occupy the continuum. Thus, halos are not easy to be formed in
the Zr isotopes.

\begin{figure}[ht!]
\includegraphics[width=0.45\textwidth]{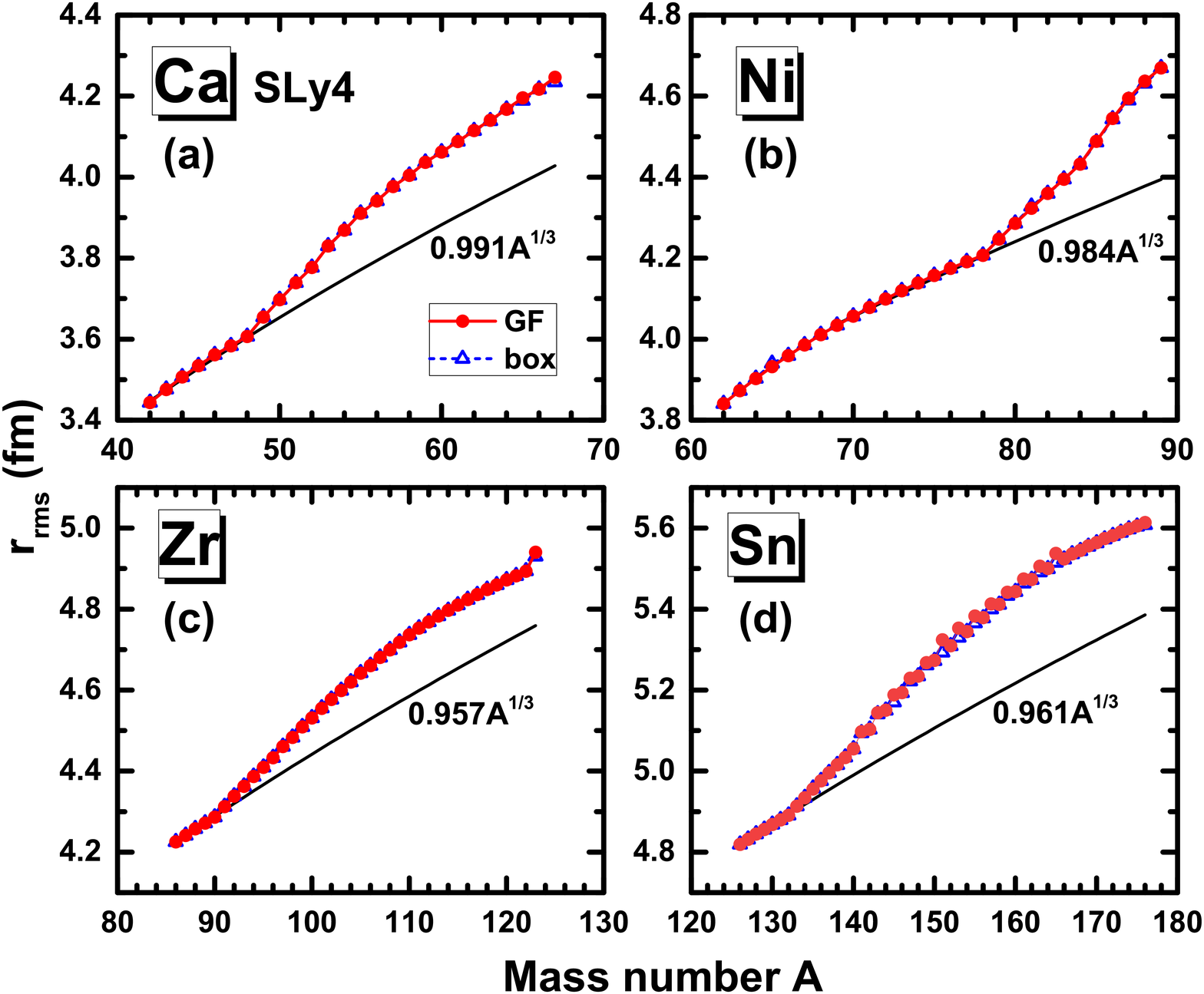}
\caption{(Color online) Neutron root-mean-square~(rms)~radii $r_{\rm rms}$ for
the (a)~Ca, (b)~Ni, (c)~Zr, and (d)~Sn isotopes calculated by the Skyrme-HFB
theory with the Green's function method~(filled circles and solid lines), in
comparison with the results by the discretized method~(open triangles and
dashed lines).} \label{Fig5}
\end{figure}

In Fig.~\ref{Fig5}, the neutron root-mean-square~(rms)~radii,
\begin{equation}
r_{\rm rms}=\sqrt{\frac{\int dr 4\pi r^{4}\rho(r)}{\int dr 4\pi r^{2}\rho(r)}},%
\label{eq:rrms}
\end{equation}%
are plotted in the Ca, Ni, Zr, and Sn isotopes, which are obtained by the
Skyrme-HFB calculations both with the Green's function method~(filled circles
and solid lines) and box-discretized method~(open triangles and dashed lines)
employing the SLy4 parameter set, in comparison with the radii $r=b_0 A^{1/3}$
in the traditional liquid-drop model (black lines), where the coefficient
$b_0$ could be determined by the radii of deeply bound nuclei. In the Ca, Ni,
Zr, and Sn chains, they are $r\thickapprox 0.991A^{1/3}$, $0.984A^{1/3}$,
$0.957A^{1/3}$, and $0.961 A^{1/3}$, respectively decided by the rms radii of
stable nuclei with neutron number less than the shell closure $N=28, 50, 50$
and $80$. In each chain, by adding more neutrons, the nuclear rms radii
$r_{\rm rms}$ increase steeply and veer off the radii $A^{1/3}$ rule. For
example, in the Ca chain, compared with the isotopic trend in $N\leqslant 20$,
which gives an extrapolation as $r\thickapprox 0.991A^{1/3}$, the neutron rms
radii in $^{50}$Ca and the heavier isotopes display steep increases with $N$.
In those mass area, possible neutron halos may occur. Besides, obvious
odd-even staggering of rms radii could be observed in the Sn chain, where the
odd-$A$ nuclei $^{151-165}$Sn own larger rms radii than the neighbouring two
even-even nuclei. The detailed explanations could be seen in
Ref.~\cite{TTSun_PRC_2019}. The odd-even staggering phenomena in nuclear radii
and nuclear mass have attracted great interests in the past years and lots of
efforts have been done, such as the works in
Refs.~\cite{Satula_PRL_1998,Dobaczewski_PRC_2001,Litvinov_PRL_2005,Hagino_PRC_2012,Wang_PRC_2013,Coraggio_PRC_2013,Chen_PRC_2015}.

\begin{figure}[t!]
\includegraphics[width=0.45\textwidth]{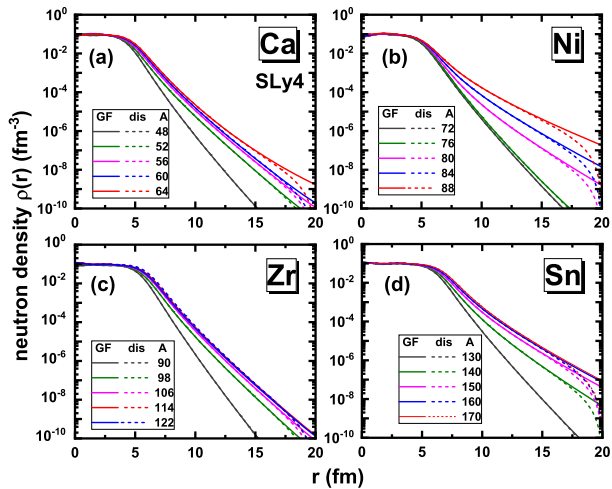}
\caption{(Color online) Neutron density $\rho(r)$ for the (a)~Ca, (b)~Ni,
(c)~Zr, and (d)~Sn isotopes calculated by the Skyrme-HFB theory with the
Green's function method~(solid lines), in comparison with the results by the
discretized method~(dashed lines).}
\label{Fig6}
\end{figure}

In the exotic nuclei, extended density distributions in the coordinate space
is often behaved. Thus, in Fig.~\ref{Fig6}, to explore the exotic structures
in the (a)~Ca, (b)~Ni, (c)~Zr, and (d)~Sn isotopes, we also plotted the
neutron density distributions $\rho(r)$, where the solid lines are those
obtained by the Green's function method in comparison with the results by the
box-discretized method denoted by the dashed lines. It is a global trend that
the neutron density distributions are extended further with the increasing
neutron number. The shell structures influence the density distribution
significantly, i.e., compared with the bound nuclei, the density distributions
of the neutron-rich nuclei in the (a)~Ca, (b)~Ni, (c)~Zr, and (d)~Sn isotopes
with the neutron number exceeding the neutron closure $N=28, 50, 50, 82$ are
much more extended, which is consistent with the behaviors of the rms radii
plotted in Fig.~\ref{Fig5}. In further, compared with the Ca and Zr chains,
the Ni and Sn chains perform the more dispersion distributions in the
densities, which can be explained by their small two-neutron separation
energies $S_{\rm 2n}$ in a large mass range plotted in Fig.~\ref{Fig1}
indicating them the very weakly bound systems. In the case of the Zr isotopes,
the density distributions in the total chain are relatively localized, and in
combination with the large $S_{\rm 2n}$ plotted in Fig.~\ref{Fig1}, we are
inclined to believe the absence of the halos in Zr isotopes. However, taking
the RCHB theory with the NLSH parameter set~\cite{JMeng_PRL_1998} and the
continuum Skyrme-HFB theory with the SKI4 parameter
set~\cite{Grasso_PRC_2006,YZhang_PRC_2012}, giant halos have been predicated.
In all the isotopes, compared with the box-discretized method which gives
nonphysical sharp decreases of the density distributions at the space
boundary, Green's function method can describe well the extended density
distributions, especially for the very neutron-rich isotopes. Besides, the
densities obtained by the Green's function method can be independent on the
space sizes as discussed in Ref.~\cite{TTSun_PRC_2019}, which is determined
basically by the proper boundary conditions of the bound states, weakly bound
sates and the continuum employed when constructing the Green's functions.

\begin{figure}[t!]
\includegraphics[width=0.45\textwidth]{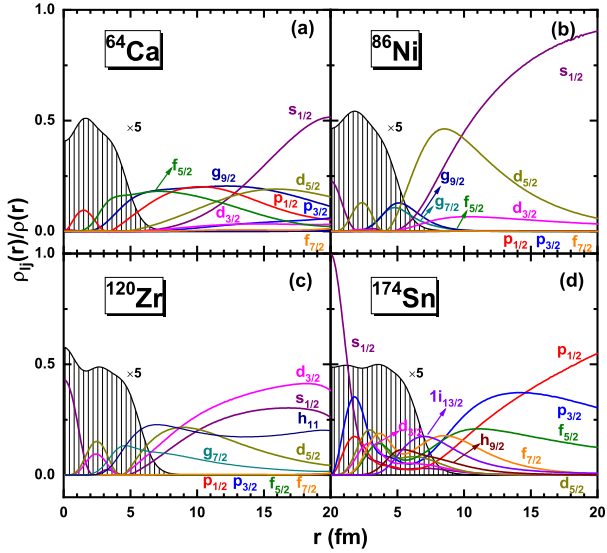}
\caption{(Color online) Contributions of different partial waves to the total
neutron density, $\rho_{lj}(r)/\rho(r)$, as a function of the radial
coordinate $r$ for the neutron-rich nuclei (a)~$^{64}$Ca, (b)~$^{86}$Ni, (c)~$^{120}$Zr, and
(d)~$^{174}$Sn. The shallow regions are for the nuclear density distributions $\rho(r)$,
which are rescaled by multiplying a factor of $5$. }%
\label{Fig7}%
\end{figure}%

\begin{figure}[t!]
\includegraphics[width=0.45\textwidth]{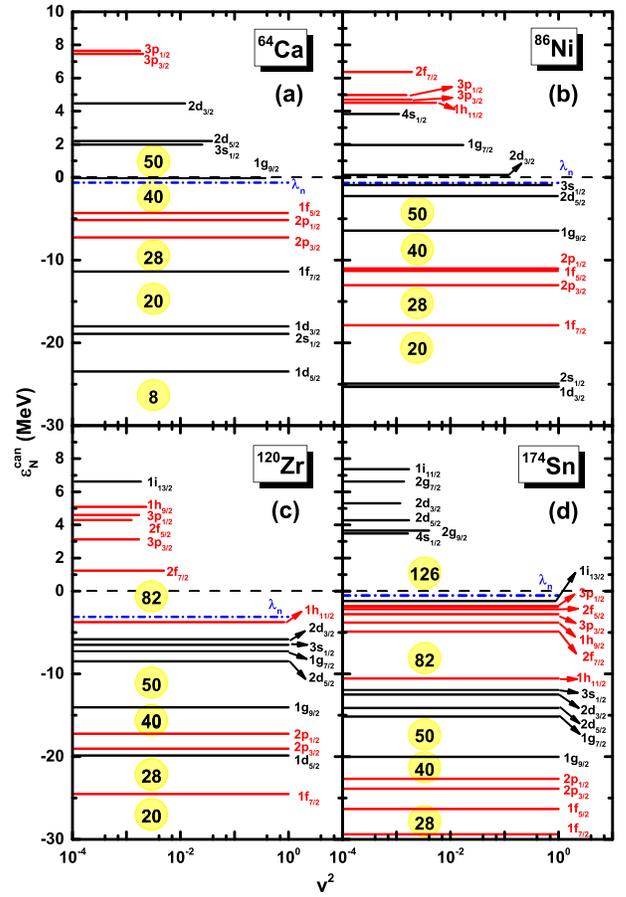}
\caption{(Color online) Neutron canonical single-particle levels in the (a)~$^{64}$Ca,
(b)~$^{86}$Ni, (c)~$^{120}$Zr, and (d)~$^{174}$Sn calculated by the Skyrme-HFB
theory with the Green's function method. The blue dashed lines represent the Fermi
surface. The occupation probabilities $v^2$ for the canonical orbitals are proportional to
the lengths of the lines (in logarithmic scale).}
\label{Fig8}
\end{figure}

To explore the contributions of different partial waves on the extended
density distributions in Fig.~\ref{Fig6}, taking the neutron-rich
(a)~$^{64}$Ca, (b)~$^{86}$Ni, (c)~$^{120}$Zr, and (d)~$^{174}$Sn as examples,
we plot in Fig.~\ref{Fig7} the compositions $\rho_{lj}(r)/\rho(r)$ as
functions of the radial coordinate $r$ and show in Fig.~\ref{Fig8} the neutron
canonical single-particle levels as well as the occupation probabilities. It
can be clearly seen that outside the nuclear surface referring to the right
boundary of the shallow regions for the total nuclear density distributions,
it's the orbitals locating around the Fermi surface that play the main role of
the density distributions. For example, in the neutron-rich $^{64}$Ca, the
partial waves $p_{1/2}$, $f_{5/2}$, $g_{9/2}$, $s_{1/2}$, and $d_{5/2}$
contribute a lot for the total density in the area of $5~{\rm fm}<r<15~{\rm
fm}$. Those levels are located around within $5~$MeV above and below the Fermi
surface as shown in Fig.~\ref{Fig8}. When going further in the coordinate
space with $r>15~$fm, the contributions of the partial waves $s_{1/2}$ and
$d_{5/2}$ with low angular momenta increase obviously while others reduce
their contributions. In the case of $^{86}$Ni, the single-particle levels
$2d_{5/2}$, $3s_{1/2}$, and $2d_{3/2}$, which are located above the neutron
shell of $N=50$ and close to the Fermi surface, play the main role of the
neutron density distribution. Although the single-particle resonant state
$1g_{7/2}$ in the continuum is also very close to the Fermi surface, very
small contribution for the density in large coordinate space is observed due
to the effect of the large centrifugal barrier. For the nucleus $^{120}$Zr
with the neutron number very close to the closure of $N=82$, the
single-particle levels between the closures $N=50$ and $N=82$ including
$2d_{5/2}$, $1g_{7/2}$, $3s_{1/2}$, $2d_{3/2}$, and $1h_{11/2}$, contribute a
lot for the densities in large coordinate space. The very small occupation of
the single-particle resonant state $2f_{7/2}$ leads to little contribution to
the density. Regarding to the neutron-rich $^{174}$Sn with the neutron number
exceeding the closure of $N=82$, the weakly bound single-particle levels
$2f_{7/2}$, $2f_{5/2}$, $3p_{3/2}$, and $3p_{1/2}$ with small angular momentum
play the key role for the extended density distributions in the large
coordinate space. According to those analysis, we can conclude that it is the
single-particle levels around the Fermi surface especially the waves with low
angular momenta that cause the extended nuclear density.

In Fig.~\ref{Fig8}, the particle occupation probabilities $v^2$ on different
canonical levels $\varepsilon_{\rm N}^{\rm can}$ are presented and denoted by
the length of the lines. Without pairing, the values of the occupation
probabilities $v^2$ should be either one or zero with the boundary line of the
Fermi surface. With the effect of pairing, the nucleons occupied the levels
below the Fermi surface can be scattered to higher levels and result in the
occupations of the weakly bound states above the Fermi surface and even levels
in the continuum. In the neutron-rich nuclei $^{64}$Ca, $^{86}$Ni, $^{120}$Zr,
and $^{174}$Sn, the number of neutrons ${\displaystyle
N_{\lambda}=\sum_{\varepsilon_k>\lambda_{\rm n}}(2j+1)v_{k}^2}$ scattered
above the Fermi surface $\lambda_{\rm n}$ are $4.33$, $2.51$, $0.159$, and
$0.173$, respectively. In the case of $^{64}$Ca, the very weakly bound
single-particle $1g_{9/2}$ contribute around $3.7$ neutrons.

\begin{figure}[t!]
\includegraphics[width=0.45\textwidth]{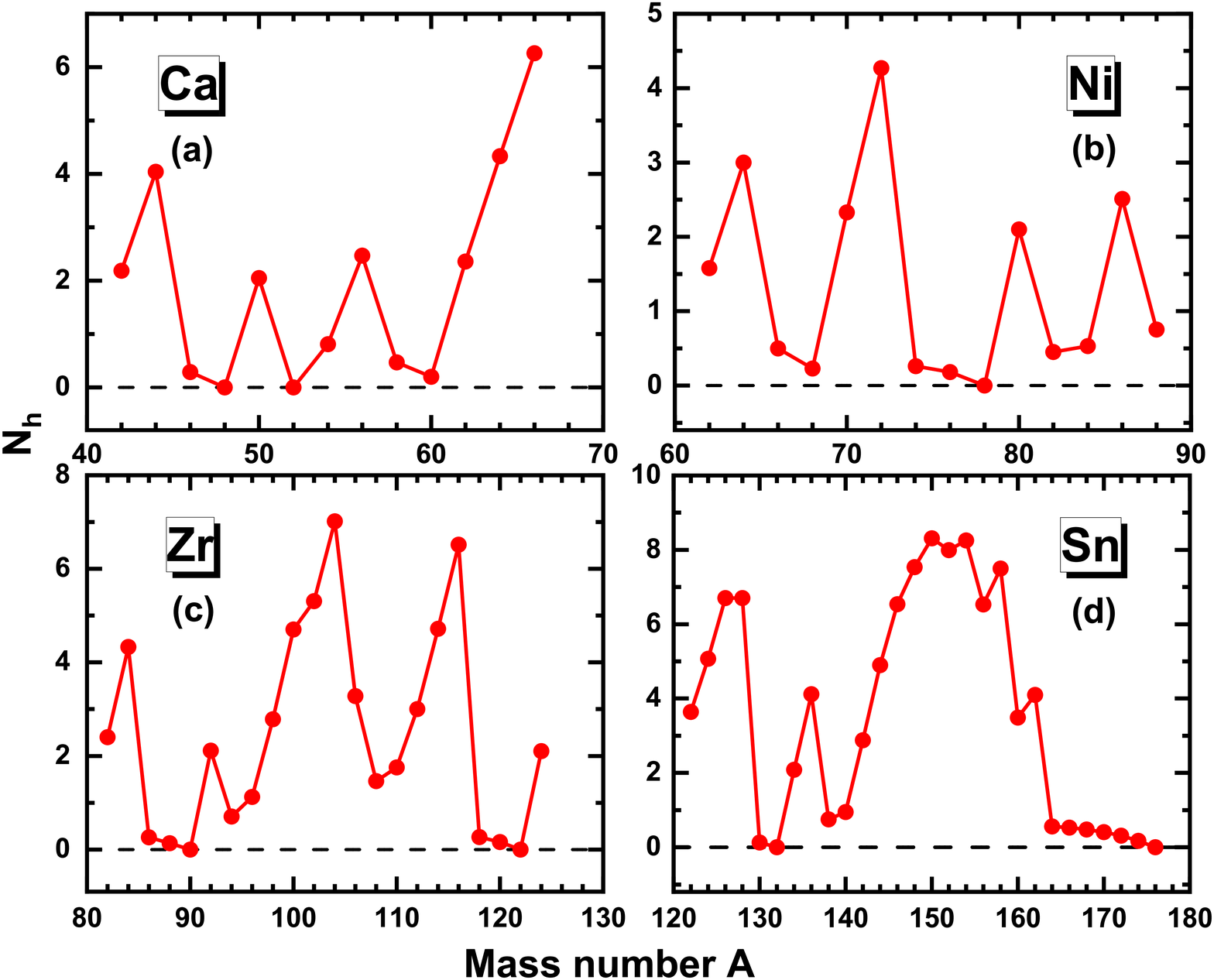}
\caption{(Color online) Number of the neutrons $N_{\lambda}$ occupied the single-particle levels above the Fermi
surface $\lambda_n$ as a function of the mass number $A$ for the (a)~Ca, (b)~Ni, (c)~Zr,
and (d)~Sn isotopes calculated by the Skyrme-HFB theory with the Green's
function method.}
\label{Fig9}
\end{figure}

To explore the effects of pairing, we plot in Fig.~\ref{Fig9} the number of
neutrons $N_{\lambda}$ scattered above the Fermi surface for the (a)~Ca,
(b)~Ni, (c)~Zr, and (d)~Sn chains obtained by the Skyrme-HFB theory with the
Green's function method. It can be seen clearly that substantial neutrons have
been scattered from the single-particle levels below the Fermi surface to the
weakly bound states above the Fermi surface and even levels in the continuum
due to the pairing, especially in the nuclei with the neutron number filling
the half-full shells. Besides, very obvious shell structure could be observed.
With the number of neutrons reaching a magic number, i.e., $N=28, 40$ in the
Ca isotopes, $N=40, 50$ in the Ni isotopes, $N=50, 82$ in the Zr isotopes, and
$N=82,126$ in the Sn isotopes, $N_{\lambda}$ is almost zero due to the absence
of the pairing for the closed-shell nuclei. Besides, at the points of $N=32$
in the Ca isotopes, $N=54,68$ in the Zr isotopes, and $N=88$ in the Sn
isotopes, very small numbers of neutron $N_{\lambda}$ are obtained, which
indicate weak pairing in those nuclei. On the other hand, it can also predict
the possible existence of sub-shells and new magic numbers. Furthermore, the
evolution of $N_{\lambda}$ is basically consistent with the trend of pair
energy $-E_{\rm pair}$ in Fig.~\ref{Fig3}.

\begin{figure}[t!]
\includegraphics[width=0.45\textwidth]{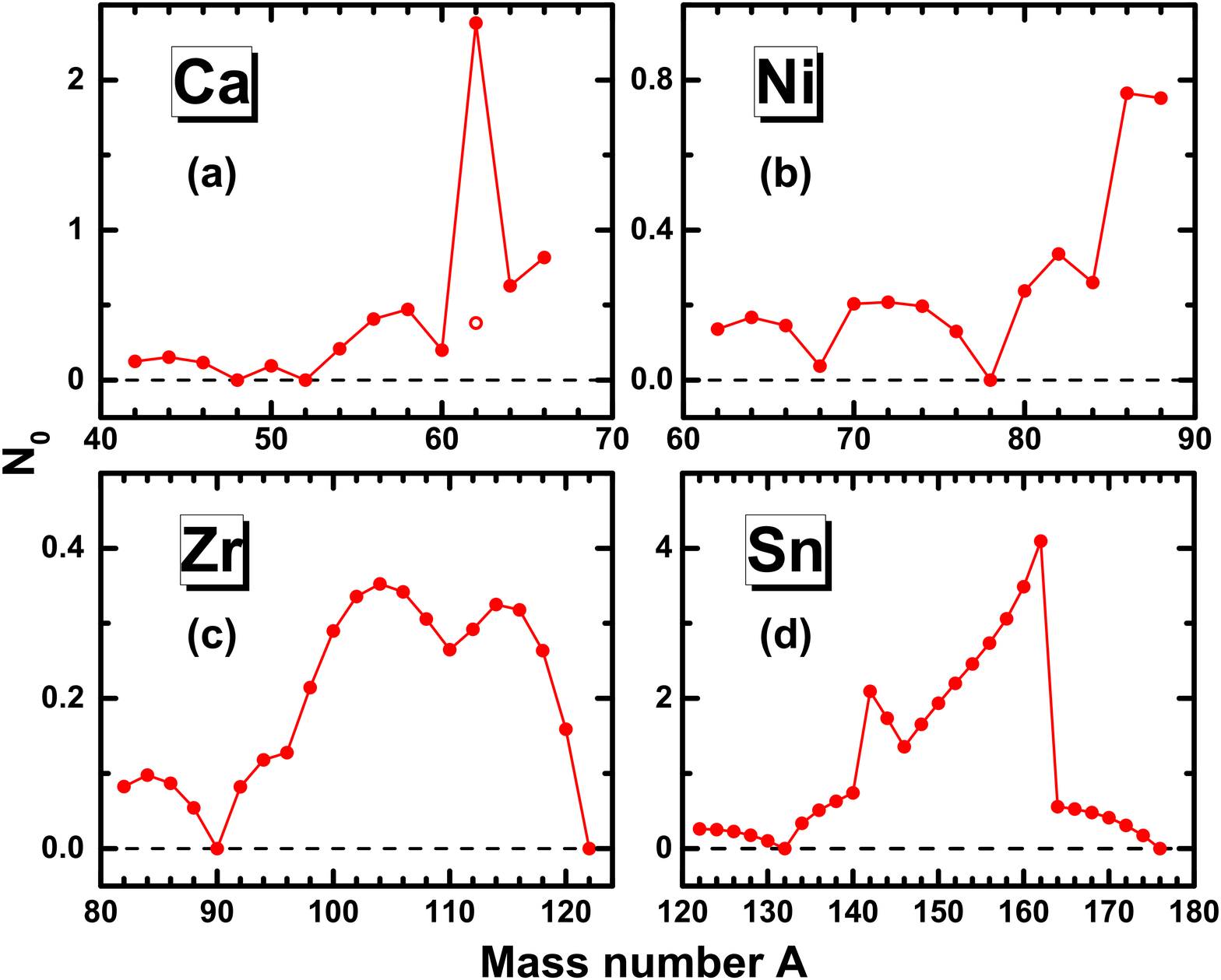}
\caption{(Color online) Number of neutrons $N_0$ occupied in the continuum (above the threshold $\varepsilon=0$~MeV)
as a function of the mass number $A$ for the (a)~Ca, (b)~Ni, (c)~Zr, and
(d)~Sn isotopes calculated by the Skyrme-HFB theory with the Green's function
method.}
\label{Fig10}
\end{figure}

In Fig.~\ref{Fig10}, we further investigate the number of neutrons occupied in
the continuum with the single-particle energies $\varepsilon
>0$~MeV, i.e., ${\displaystyle N_{0}=\sum_{\varepsilon_k> 0}(2j+1)v_{k}^2}$. Compared
with the number of neutron $N_{\lambda}$ occupied in the levels above the
Fermi surface, the number of neutrons occupied in the continuum $N_0$ is
reduced drastically. For example, in the Ca chain, $N_{0}$ is less than one in
all isotopes except $^{62}$Ca. In $^{62}$Ca, the single-particle level
$1g_{g/2}$ displays as a low-lying resonant state with energy of
$\varepsilon=0.198$~MeV and a high occupation probability of $v^2 = 0.197$,
resulting to almost $1.979$ neutrons occupying on it. But in the neighboring
$^{60}$Ca, very small occupation probability $v^2=0.015$ of $1g_{9/2}$ is
obtained while in $^{64}$Ca, the $1g_{9/2}$ state drops to be a weakly bound
level with the energy of $\varepsilon=-0.072$~MeV. After removing the neutron
contribution from $1g_{g/2}$ in $^{62}$Ca, only $0.38$ neutrons are in the
continuum, which is denoted by an empty circle in panel (a). Except the Sn
chain, the shape of $N_0$ is very close to those of pairing energy and pairing
gap. Although the case of Sn chain becomes very complex, we can still observe
the shell structure at $N=82$ and $N=126$ where $N_0$ is almost zero.

\section{SUMMARY}
\label{Sec:Summary}

In this work, the exotic nuclear properties of neutron-rich Ca, Ni, Zr, and Sn
isotopes are studied systematically taking the continuum Skyrme-HFB theory in
the coordinate space formulated with the Green's function method, in which the
pairing correlations, the couplings with the continuum, and the blocking
effects for the unpaired nucleon in odd-$A$ nuclei are treated properly.

Firstly, both the two-neutron separation energies $S_{\rm 2n}$ and one-neutron
separation energies $S_{\rm n}$ are calculated, which consist well with the
experimental data. Great differences exist for the drip lines determined by
$S_{\rm 2n}$ and $S_{\rm n}$, and in the Ca, Ni, Zr, and Sn isotopes, the
determined drip line nuclei are $^{67}$Ca, $^{89}$Ni, $^{123}$Zr, and
$^{177}$Sn judging by $S_{\rm 2n}$ while they are $^{60}$Ca, $^{86}$Ni,
$^{122}$Zr, and $^{148}$Sn according to $S_{\rm n}$. For further research, the
pairing energies for even-even and odd-even nuclei are studied. We found that
the neutron pairing energies $-E_{\rm pair}$ of the odd-$A$ nuclei are around
$2$~MeV smaller in comparison with those of the neighboring even-even nuclei,
which can be explained by the absent contribution of pairing energy by the
single unpaired odd neutron. This also explains why the drip lines determined
by $S_{\rm n}$ are much shorter than those by $S_{\rm 2n}$. In addition, from
the fluctuation trends of the pairing energy, the traditional neutron magic
numbers are clearly displayed, i.e., $N=28, 40$ in the Ca isotopes, $N=40, 50$
in the Ni isotopes, $N=50, 82$ in the Zr isotopes, and $N=82,126$ in the Sn
isotopes.

Secondly, to explore the possible halo structures in the neutron-rich Ca, Ni,
Zr, and Sn isotopes, the neutron single-particle structures, the
root-mean-square radii, and the density distributions are investigated. In the
neutron-rich Ca, Ni, Sn nuclei, especially the weakly bound nuclei close to
neutron drip line, the rms radii have sharp increases with significant
deviations from the traditional $r\varpropto A^{1/3}$ rule. Besides, very
diffuse spatial density distributions are also observed in those nuclei, which
reflects the possible halo phenomenon therein. By analyzing the contribution
of different partial waves to the total density, we found that it is the
orbitals locating around the Fermi surface especially the waves with low
angular momenta that cause the extended nuclear density and large rms radii.

Finally, the number of halo nucleons which could reflects the effects of
pairing are discussed. Two different types of the number of the neutrons are
defined, i.e., $N_{\lambda}$ occupied the single-particle levels above the
Fermi surface $\lambda_n$ and $N_0$ occupied in the continuum. We found that
the evolutions of $N_{\lambda}$ and $N_0$ with the mass number $A$ are
basically consistent with the trend of pairing energy $-E_{\rm pair}$, which
supports the key role of the pairing correlations in the halo phenomena.

\begin{acknowledgements}
This work was partly supported by the National Natural Science Foundation of
China~(U2032141), the Natural Science Foundation of Henan
Province~(202300410479), the Foundation of Fundamental Research for Young
Teachers of Zhengzhou University (JC202041041), and the Physics Research and
Development Program of Zhengzhou University (32410217). The theoretical
calculation was supported by the nuclear data storage system in Zhengzhou
University.
\end{acknowledgements}

%

\end{document}